# Coherent perfect nanoabsorbers based on negative refraction

Vasily Klimov[1]*, Shulin Sun[2,3], Guang-Yu Guo[2,4]♣

[1]*P. N. Lebedev Physical Institute, Russian Academy of Sciences, 119991 Moscow, Russia*
[2]*Department of Physics, National Taiwan University, Taipei 10617, Taiwan*
[3]*National Center of Theoretical Sciences at Taipei, Physics Division, National Taiwan University, Taipei 10617, Taiwan*
[4]*Graduate Institute of Applied Physics, National Chengchi University, Taipei 11605, Taiwan*
*[vklim@sci.lebedev.ru](mailto:vklim@sci.lebedev.ru)*; ♣*[gyguo@phys.ntu.edu.tw](mailto:gyguo@phys.ntu.edu.tw)*

**Abstract:** Based on both analytical dipole model analyses and numerical simulations, we propose a concept of coherent perfect nanoabsorbers (CPNAs) for divergent beams. This concept makes use of the properties of a slab with negative refraction and small losses. The proposed CPNA device would allow focusing radiation in nanoscale regions, and hence could be applied in optical nanodevices for such diverse purposes as reading the results of quantum computation which is based on single photon qubits.

# 1. Introduction

Recently, a fascinating concept of coherent perfect absorbers (CPAs) has been suggested [1, 2]. In some sense this concept is an inverse to the concept of lasing. This conception was demonstrated experimentally by the example of a symmetric Fabry-Perot cavity with small losses which was symmetrically irradiated by 2 coherent plane waves (see Fig. 1). In [1, 2], it was shown that under certain conditions, all the energy of incoming waves is fully absorbed by the dielectric (silicon) slab. As a result, the radiation from the system disappears for some parameters. This situation indeed is reciprocal to the case of lasing, where for high enough pumping the coherent radiation appears. More complicated examples of CPA with plane waves were considered in [3-5].

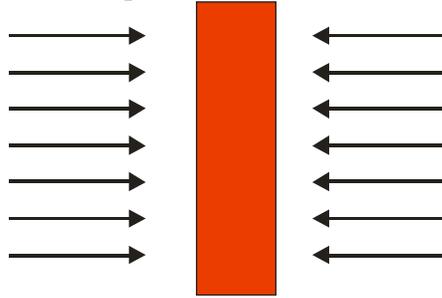

Fig. 1. Schematic diagram of a plane wave coherent perfect absorber.

Despite generic nature of this concept, its application to more complicated geometries is a nontrivial task. In particular, since now the control of radiation of atoms and molecules with nanoparticles and metamaterials becomes very important, it is desirable to have an effective nanoabsorber of electromagnetic fields from a single atom or molecule. First of all, one may have a complicated spatial structure of light field. For example, the emission pattern of an atom would correspond to spherically divergent beams rather than plane waves. Moreover, many applications nowadays are related with making use of metamaterials for the control of light at the nanoscale [6, 7].

In the present work, we propose a concept of coherent perfect nanoabsorbers (CPNAs) using a slab made of double negative (DNG) metamaterial, that is, the metamaterial with negative refractive index [6]. Nowadays, slabs with negative refraction are widely used for many applications such as perfect lensing [8] and cloaking of small objects [9]. Our concept is based on recently discovered focusing properties of a system of sources and sinks near a slab with negative refraction [10, 11], where it was shown that if two sources and one sink were put at the ray intersection points, all the energy from the sources would go to the point sink. In this paper we propose to use a real nanoparticle instead of the hypothetic sink used in [10,11]. The operation scheme of our CPNA is shown schematically in Fig. 2. The device shown in Fig. 2 would allow absorbing all the energy from the 2 sources emitted in the direction towards the negative refraction slab by a nanoparticle placed inside it.

It should be emphasized that within the geometry of Fig. 2, only half of the radiated energy is going towards the slab and then absorbed. However, it is a trivial task to slightly modify this system by adding perfect magnetic mirrors just behind the $x$-polarized dipoles. This would result in doubling of the radiated energy going to the slab and then absorbed. For $z$-oriented dipoles, one should use perfect electric mirrors to absorb all the radiated energy. For simplicity, we will consider here only the systems without mirrors.

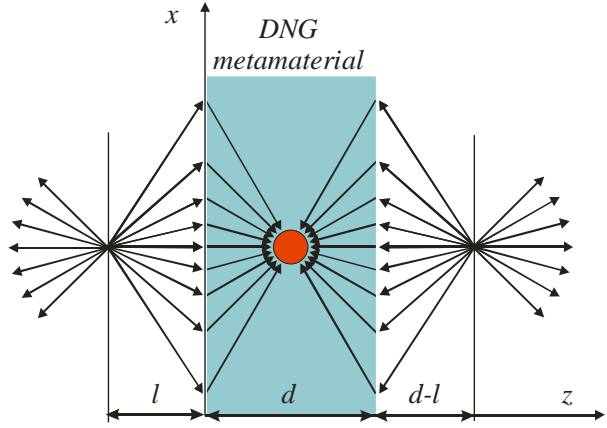

Fig. 2. Schematic diagram of a coherent perfect nanoabsorber for the divergent beams. All the energy from the 2 point sources is absorbed by the nanoparticle (red circle), placed inside the negative refraction slab.

## 2. Analytical formulation

To describe the idea of our proposal, let us start from a 2D DNG slab without losses with $\varepsilon_2 = -1$, $\mu_2 = -1$ placed in vacuum with $\varepsilon_1 = 1$, $\mu_1 = 1$. Let us put into this system 3 2D dipoles which are described by line currents

$$j_{1,x} = -i\omega p_0 \delta(x)\delta(z+l), \quad j_{2,x} = -i\omega p_0 \delta(x)\delta(z+l-2d); \quad (1)$$
$$j_{3,x} = +i\omega p_0 \delta(x)\delta(z-l)$$

where $p_0$ is the dipole moment of the wires per unit length, and everywhere the time dependence $e^{-i\omega t}$ is assumed. It can be easily verified that the magnetic fields from these currents will have the following form

$$H_{0,y} = k_0 \pi p_0 \frac{\partial}{\partial z} H_0^{(1)}\left(k_0\sqrt{x^2+(z+l)^2}\right), z<0$$
$$H_{0,y} = -k_0 \pi p_0 \frac{\partial}{\partial z} H_0^{(1)}\left(k_0\sqrt{x^2+(z-l)^2}\right), 0<z<d \quad (2)$$
$$H_{0,y} = k_0 \pi p_0 \frac{\partial}{\partial z} H_0^{(1)}\left(k_0\sqrt{x^2+(z-2d+l)^2}\right), z>d$$

where $k_0 = \omega/c$ and $H_0^{(1)}(x)$ is the Hankel function. The electric fields can then be found with the help of Amper's law

$$rot\mathbf{H} = -ik_0 \varepsilon \mathbf{E}. \quad (3)$$

From Eq. (2) one can see that there are 3 outgoing 2D waves for the assumed $e^{-i\omega t}$ time dependence. However, taking into account the fact that in the slab we have negative refraction, we immediately see that 2 of these waves (right and left) correspond to the energy sources while the third one, which is placed into the negative refraction slab, corresponds to an energy sink. The distribution of the energy flows for solution Eq. (2) is shown in Fig. 3, and we can see from this figure that all the energy from the 2 sources radiated in the direction of the slab is absorbed by the sink dipole.

It is easy to show this directly by calculating the work of the field over the sink dipole, that is the absorbed power,

$$W_{abs} = +\frac{\omega}{2}\operatorname{Im} p_0 \mathbf{E}^*(\mathbf{r}_{abs}) = \frac{\omega}{2}k_0^2|p_0|^2\frac{\pi}{2}. \qquad (4)$$

Because there is no reflection, the expressions for the fields are the same in all regions [see Eq.(2)]. It is easy to see that the absorbed energy Eq. (4) is equal to the total energy radiated by one source or half of the energy radiated by 2 sources

$$W_{rad} = -\frac{\omega}{2}\operatorname{Im} \mathbf{p}_0 \mathbf{E}^*(\mathbf{r}_1) = \frac{\omega}{2}k_0^2|p_0|^2\frac{\pi}{2}. \qquad (5)$$

Thus, the 3 dipole system Eq. (2) is a perfect realization of our CPNA if one approximates the nanoparticle by a dipole.

However, real systems cannot be without losses and it is important to demonstrate the idea of coherent perfect nanoabsorber in more realistic situation with losses. To do this, let us consider a DNG slab with losses, that is $\varepsilon_2 = -1+i\delta$, $\mu_2 = -1+i\delta$, and again place into this system 3 dipoles. However, now we do not know what amplitude of the dipole moment of the nanoabsorber would result in the CPNA, because now we should take into account its self-consistent interaction with the slab and other sources. Therefore, now we will consider a system of currents

$$j_{1,x} = -i\omega p_0 \delta(x)\delta(z+l), \; j_{2,x} = -i\omega p_0 \delta(x)\delta(z+l-2d);$$
$$j_{3,x} = -i\omega p_{abs}\delta(x)\delta(z-l) \qquad (6)$$

where the amplitude of the dipole moment, $p_{abs}$, inside the slab is an unknown quantity to be determined through the condition of perfect absorption. To prove the existence of CPNA for this case, let us again calculate the energy absorbed by the dipole inside the slab

$$W_{abs} = +\frac{\omega}{2}\operatorname{Im} \mathbf{p}_{abs}\mathbf{E}^*(\mathbf{r}_{abs}). \tag{7}$$

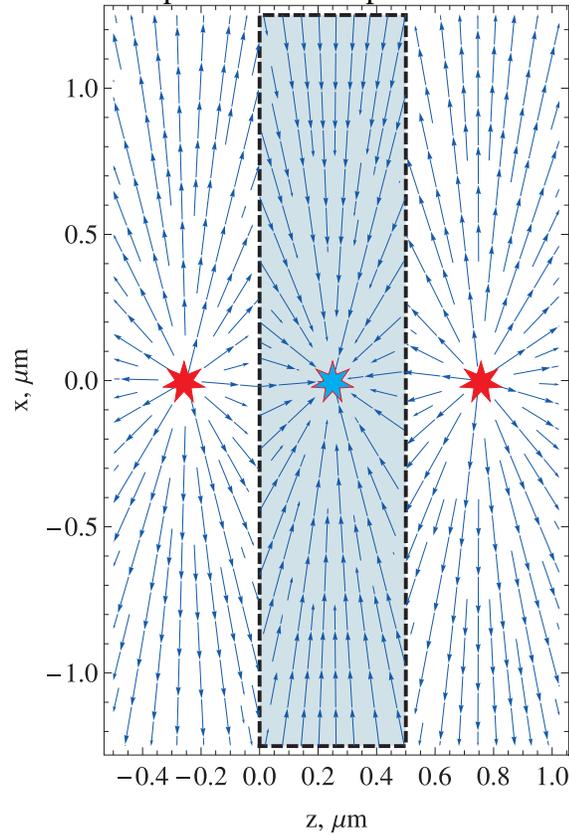

Fig. 3. Energy flows from the left and right sources to the sink in the middle of the slab without losses with $\varepsilon_2 = -1$, $\mu_2 = -1$ (light blue rectangle) placed in vacuum with $\varepsilon_1 = 1$, $\mu_1 = 1$ [according to solution (2)].

Now, however, the fields are more complicated. In particular, the field near this absorbing dipole is composed also by reflections from the walls and from the dipoles placed outside the slab, that is

$$\mathbf{E}(\mathbf{r}_{abs}) = \mathbf{E}^{(abs),0}(\mathbf{r}_{abs}) + \mathbf{E}^{(abs),R}(\mathbf{r}_{abs}) + \mathbf{E}^{(1)}(\mathbf{r}_{abs}) + \mathbf{E}^{(2)}(\mathbf{r}_{abs}) \tag{8}$$

where $\mathbf{E}^{(1)}(\mathbf{r}_{abs})$ and $\mathbf{E}^{(2)}(\mathbf{r}_{abs})$ are, respectively, the fields from first and second sources at the center of the absorbing nanoparticle, while $\mathbf{E}^{(abs),0}(\mathbf{r}_{abs})$ and $\mathbf{E}^{(abs),R}(\mathbf{r}_{abs})$ are, respectively, the fields from $\mathbf{p}_{abs}$ in unbound space with $\varepsilon_2$ and $\mu_2$ and corresponding reflected part at the center of the absorbing nanoparticle. If one presents electric fields through Green functions of the dipole source $\mathbf{p_0}$ placed at $\mathbf{r_0}$, that is through relation

$$\mathbf{E}(\mathbf{r}) = \vec{G}(\mathbf{r},\mathbf{r}_0)\mathbf{p}_0; \quad \vec{G}(\mathbf{r},\mathbf{r}_0) = \vec{G}^{(0)}(\mathbf{r},\mathbf{r}_0) + \vec{G}^{(R)}(\mathbf{r},\mathbf{r}_0), \tag{9}$$

one obtains

$$W_{abs} = -\frac{\omega}{2}\operatorname{Im} p^*_{abs}\left(G^{(abs)0}_{xx}p_{abs} + G^{(abs)R}_{xx}p_{abs} + G^{(1)}_{xx}p_0 + G^{(2)}_{xx}p_0\right) \tag{10}$$

instead of Eq. (7). In Eq. (10), we have taken into account the fact that the fields on the z-axis have only nonzero x-component. Eq. (10) is a bilinear form of the $p_{abs}$ and it will definitely have a maximum which would provide the perfect absorption. To find this maximum, we should find all the Green functions entering Eq. (10).

First of all, let us find the self-action part of the field, $G^{(abs)0}_{xx}$. It can be easily found because this part appears as a solution in uniform space with parameters of the slab, that is, in this case

$$\begin{aligned}H_{0,y} &= k_0\pi p_{abs}\frac{\partial}{\partial z}H^{(1)}_0\left(k_2\sqrt{x^2+(z-l)^2}\right) \\ E_{0,x} &= -\frac{i\pi p_{abs}}{\varepsilon_2}\frac{\partial^2}{\partial z^2}H^{(1)}_0\left(k_2\sqrt{x^2+(z-l)^2}\right) = G^{(abs)0}_{xx}p_{abs}\end{aligned} \tag{11}$$

where $k_2 = \sqrt{\varepsilon_2\mu_2}$. Thus, at the absorber position, we have

$$G^{(abs)0}_{xx} = -\frac{i\pi}{\varepsilon_2}\frac{\partial^2}{\partial z^2}H^{(1)}_0\left(k_2|z-l|\right)\bigg|_{z=l}. \tag{12}$$

To calculate Eq. (12), one should use an integral presentation for Hankel function [12]

$$H^{(1)}_0\left(k_2\sqrt{x^2+(z)^2}\right) = \frac{1}{\pi}\int_{-\infty}^{+\infty}dq\, e^{iqx}\frac{e^{ik_{z,2}|z|}}{k_{z,2}}. \tag{13}$$

In Eq. (13) and other places, we use notion $k_{z,1} = \sqrt{k_0^2\varepsilon_1\mu_1 - q^2}, k_{z,2} = \sqrt{k_0^2\varepsilon_2\mu_2 - q^2}$ ($\operatorname{Im} k_z > 0$) for longitudinal wavevectors and $q$ is its transversal part. As a result, the expression for free Green function at the origin will have the form

$$G^0_{xx,abs} = \frac{i}{\varepsilon_2}\int_{-1/R}^{1/R}dq\sqrt{k_0^2\varepsilon_2\mu_2 - q^2} \tag{14}$$

where integration here is in finite limits, from $-1/R$ to $1/R$, and $R$ is the radius of the nanoparticle. The necessity to limit the integration domain is related with the fact that the absorbed or radiated power of the point source is infinite even in media with arbitrary low losses.

To find other Green functions in Eq.(10), let us start from the case when the DNG slab with arbitrary dielectric constant is illuminated with line source #1 with current density

$$j_x = -i\omega p_0 \delta(x)\delta(z+l) \tag{15}$$

where $p_0$ is the dipole moment of wire per unit length. For such a source, the magnetic field in free space with $\varepsilon=1$, $\mu=1$ and the wavenumber $k=\omega/c$ has only one component which can be presented in the form of Eq. (2). Now using integral presentation for Hankel function Eq. (13), the #1 source field in unbounded space can be presented in the form

$$H_{0,y} = ik_0 p_0 \int_{-\infty}^{+\infty} dq\, sign(z+l) e^{iqx+ik_{z,1}|z+l|}. \tag{16}$$

Using this presentation, one can write general expressions for the fields in all regions of space (see Fig. 2)

$$H_y = ik_0 p \int_{-\infty}^{+\infty} dq\, e^{iqx} \left[ sign(z+l)e^{ik_{z,1}|z+l|} + A(q)e^{-ik_{z,1}z} \right], z<0$$

$$H_y = ik_0 p \int_{-\infty}^{+\infty} dq\, e^{iqx} \left[ B(q)e^{ik_{z,2}z} + C(q)e^{-ik_{z,2}z} \right], 0<z<d \tag{17}$$

$$H_y = ik_0 p \int_{-\infty}^{+\infty} dq\, e^{iqx} \left[ D(q)e^{ik_{z,1}z} \right], z>d$$

The coefficients $A(q)$, $B(q)$, $C(q)$, $D(q)$ can be found from the continuity of tangential components of magnetic and electric ($E_x = -\frac{i}{k_0 \varepsilon}\frac{\partial H_y}{\partial z}$) fields. Now finding from Eq. (17) the $x$-component of the electric field and setting $z=l$ and $x=0$, we will obtain $G_{xx}^{(1)}(\mathbf{r}_{abs})$. Analogously, one can find $G_{xx}^{(2)}(\mathbf{r}_{abs})$ and $G_{xx}^{(abs)R}(\mathbf{r}_{abs})$. In the fully symmetric case $l=d/2$ these expressions can be presented in the form

$$G_{xx}^{(1)}(\mathbf{r}_{abs}) = G_{xx}^{(2)}(\mathbf{r}_{abs}) = -2i\int dq\, k_{z,1} k_{z,2} e^{id/2(k_{z,1}+k_{z,2})} \frac{(k_{z,1}\varepsilon_2 + k_{z,2}) + (k_{z,1}\varepsilon_2 - k_{z,2})e^{ik_{z,2}d}}{(k_{z,1}\varepsilon_2 - k_{z,2})^2 e^{2ik_{z,2}d} - (k_{z,1}\varepsilon_2 + k_{z,2})^2}$$

$$G_{xx}^{abs(R)}(\mathbf{r}_{abs}) = 2i\int dq\, \frac{k_{z,2}}{\varepsilon_2} e^{ik_{zz,2}d} \frac{(k_{z,1}^2\varepsilon_2^2 - k_{z,2}^2) + (k_{z,1}\varepsilon_2 - k_{z,22})^2 e^{ik_{z,2}d}}{(k_{z,1}\varepsilon_2 - k_{z,2})^2 e^{2ik_{z,2}d} - (k_{z,1}\varepsilon_2 + k_{z,2})^2}. \tag{18}$$

Now all the Green functions for any parameters can be calculated numerically since we know their analytical expressions. In Fig. 4, we display the absorbed energy as a function of the relative amplitude of the dipole moment of the nanoparticle $\xi = p_{abs}/p_0$ (the solid curve).

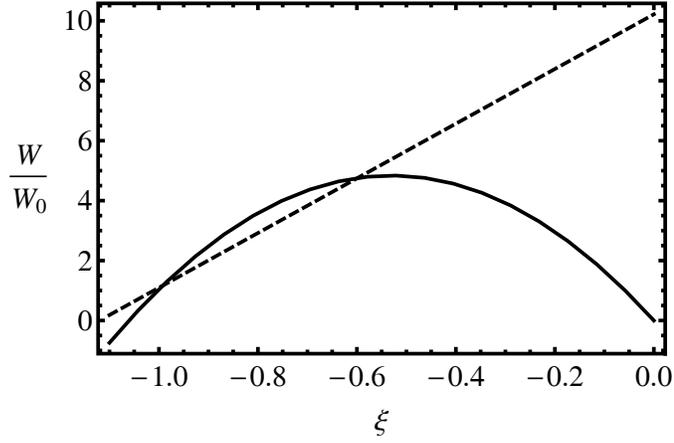

Fig. 4. The power radiated by the source ($W_{rad}$, dashed line) and the power absorbed by the nanoparticle ($W_{abs}$, solid line) relative to the power radiated by one source ($W_0$) in free space, as a function of the relative amplitude of the dipole moment of the nanoparticle $\xi = p_{abs}/p_0$. It is assumed that the slab is made of a metamaterial with $\varepsilon_2 = -1+i0.03$, $\mu_2 = -1+i0.03$, $\lambda=3$ µm, $d = 500$ nm, $l = 250$ nm.

We can see from Fig. 4 that indeed for $\xi \approx -0.6$, there is a maximum of absorption which corresponds to CPNA. Here it is important to note that the energy absorbed by the nanoparticle is 5 times greater than the power of the single source in free space. Of course there is no violation of the energy conservation law here because, due to the Purcell effect [13,14], the radiated power can be substantially enhanced by the environment.

To calculate the effective emitted energy, one should again calculate the work done by the source over the field, that is

$$W_{rad} = -\frac{\omega}{2}\operatorname{Im}\mathbf{p}_0\mathbf{E}^*(\mathbf{r}_1) \tag{19}$$

Now, the field near the dipole of the first source is formed also by reflections from the walls and from the other dipoles

$$\mathbf{E}(\mathbf{r}_1) = \left(\mathbf{E}^{(1)0}(\mathbf{r}_1) + \mathbf{E}^{(1)R}(\mathbf{r}_1) + \mathbf{E}^{(2)}(\mathbf{r}_1) + \mathbf{E}^{(abs)}(\mathbf{r}_1)\right) \tag{20}$$

where $\mathbf{E}^{(1)0}(\mathbf{r}_1), \mathbf{E}^{(1)R}(\mathbf{r}_1)$ and $\mathbf{E}^{(2)}(\mathbf{r}_1)$ are fields from first and second sources at the first source position $\mathbf{r}_1$, while $\mathbf{E}^{(abs)}(\mathbf{r}_1)$ is the field from the nanoparticle dipole at the first source position. If one again present the electric fields through Green functions [Eq. (9)], one obtains

$$W_{rad} = +\frac{\omega}{2}\operatorname{Im}p_0^*\left(G_{xx}^{(1)0}(\mathbf{r}_1)p_0 + G_{xx}^{(1)R}(\mathbf{r}_1)p_0 + G_{xx}^{(2)}(\mathbf{r}_1)p_0 + G_{xx}^{(absorber)}(\mathbf{r}_1)p_{abs}\right) \tag{21}$$

instead of Eq. (19). Green functions in Eq. (21) can be calculated in full analogy with the calculation of the Green functions of Eq. (18). The dashed line in Fig. 4 shows the dependence of normalized radiated energy [Eq. (21)] on the relative amplitude of the nanoparticle dipole moment. One can see that indeed the emitted energy grows linearly as the dipole moment of the absorber increases.

Another important feature of Fig.4 is that for relative dipole amplitude $\xi > 0$ and for $\xi < \approx -1$, the absorption become negative such that it changes to radiation, that is, some kind of lasing occurs if we provide such dipole amplitudes. However, as we will see below, it is not an easy task because these conditions need negative absorption in nanoparticles, which is gain media of the nanoparticle.

Since we have approximated the nanoparticle by a point dipole, the results of our theoretical calculations presented in this section could be only qualitative. To describe the CPNA in a quantitative manner, one should use full scale computer simulations. Nevertheless, the analytic dipole model analyses reported here give us a clear understanding of the physics underlying CPNA.

## 3. Numerical simulations

To illustrate the principle of operation of such type devices, we have also carried out numerical simulations of the system shown in Fig. 2, using the finite elements method (FEM) within the COMSOL Multiphysics software. We use a nanocylinder with parameters $\varepsilon_c, \mu_c$ as the absorbing element and 2 dipole nanowires with current densities

$$j_{1,x} = -i\omega p_0 \delta(x)\delta(z+l), j_{2,x} = -i\omega p_0 \delta(x)\delta(z+l-2d) \qquad (22)$$

as sources. That it, we are considering a 2D geometry with $p_0$ as the dipole moment of the wire per unit length and only nonzero $H_y$ component of the magnetic field.

Typical distribution of the energy flows is displayed in Fig. 5. Figure 5 shows that for the chosen parameters of the cylinder, indeed, all the energy from the sources incident on the slab is totally absorbed by the cylinder. From Fig. 5, one can see also that the energy flows in this case are very close to the results obtained for the 2 point sources and one point sink in the slab without losses (see Fig. 3).

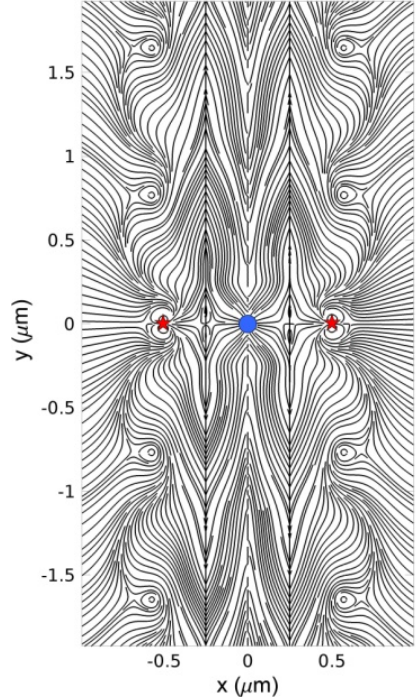

Fig. 5. Plot of energy flow lines in a coherent perfect nanoabsorber, illustrating that almost all streamlines outgoing from the sources (denoted by the red stars) to the slab are converged to the nanoparticle (denoted by the blue circle). Parameters of the simulation are $\varepsilon_2$ = -1+0.03i, $\mu_2$ = -1+0.03i, $d$ =500 nm for the slab; $\varepsilon_c$ =1.3+0.4i, $\mu_c$ =1, $R$ = 50 nm for the cylinder; and $\lambda$ =3 μm, $l$=250 nm for the sources.

In Fig. 6, the distribution of the squared electric fields in CPNA is shown. One can see from Fig. 6 that indeed the electric fields have maxima at the source positions and also at the absorbing nanoparticle position. This fact strongly supports the idea of symmetry between sources and sinks which follows from the idealized solution Eq. (2). It can be also seen from Fig. 6 that the electric fields at the interface have substantially smaller amplitudes which means that the surface plasmon excitation is indeed suppressed despite very small losses in the slab metamaterial.

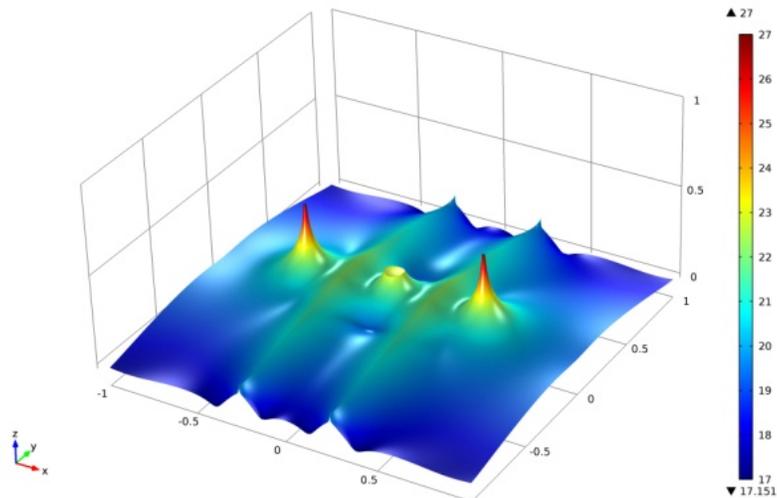

Fig. 6. Electric field intensity distribution (logarithmic scale) in the coherent perfect nanoabsorber. Simulation parameters are $d$=500nm, $l$=250nm, $\lambda$=3μm, $\varepsilon_2 = \mu_2 = -1+0.03i$, $R$=50nm, $\varepsilon_c$=1+0.21$i$, $\mu_c$=1.

To find the optimal conditions for operation of CPNA, we have displayed the energy absorbed by the nanoparticle as a function of both the real and imaginary parts of its permittivity as a 2D contour plot in Fig. 7

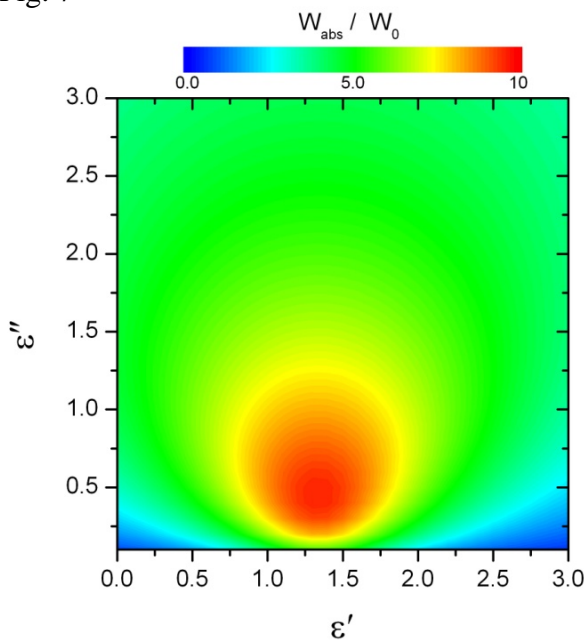

Fig. 7. Power absorbed by the nanocylinder ($W_{abs}$) normalized to the power of one source in free space ($W_0$) as a function of both the real and imaginary parts of the cylinder permittivity. Other parameters are the same as in Fig. 6

. From this figure one can see that at $\varepsilon_c = 1.3+0.4i$ there is a rather broad maximum of the absorbed power and that this maximum is about 10 times more than the power radiated by the single source in free space. To understand how this happens, we have plotted in Fig. 8 both the emitted and absorbed energy as a function of the imaginary part of the permittivity of the absorbing nanoparticle.

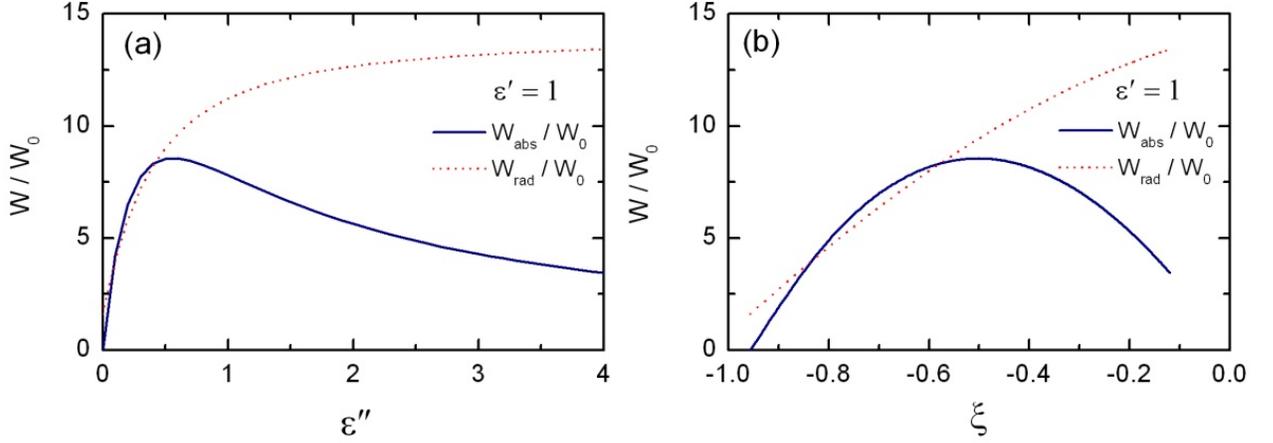

Fig.8. Radiated and absorbed powers normalized to the power of one source in free space as a function of the imaginary part of the cylinder permittivity (a) and also of the dipole moment of the cylinder $\xi = p_{cyl}/p_0$ (b). Simulation parameters are $R = 50$ nm, $d = 500$ nm, $l = 250$ nm, $\lambda = 3\mu m$, $\varepsilon_2 = \mu_2 = -1+0.03i$, $\varepsilon_c = 1+i\varepsilon''$, $\mu_c = 1$.

From Fig. 8a, one can see that the absorbed energy is even greater than the energy radiated by only one source in the region $0.1 < \varepsilon'' < 0.5$. Of course, there is no violation of energy conservation here. The reason is that due to the presence of the absorber, the radiation pattern from the sources becomes asymmetric and more than ½ of its energy is going to the slab. Note that one can increase this contribution to 100% by introducing corresponding mirrors. Another interesting feature of Fig. 8 is the dependence (growth) of energy radiated by the source on the absorber parameters. Again there is no magic. It is the well known Purcell effect [13,14] and the nanoparticle increases the radiation power of the source. Note that the feedback effects of the fields on the source current amplitude are not taken into account here. Fig. 8b shows the absorbed and radiated energies as a function of the dipole moment of the nanoparticle. It is gratifying to see that the numerical results in Fig. 8b are in good agreement with that of the analytical model analyses described in the preceding section (Fig. 4). In contrast to Fig. 8a, the dipole moment of the nanoparticle in principle can take any value. Interestingly, for $\xi > 0$ and also for $\xi <\approx -1$, the absorption becomes negative, and this corresponds to the negative values of the imaginary part of the cylinder permittivity, that is, to a gain media, as can be seen by a comparison of Fig. 8b with Fig. 8a. This, excitingly, indicates that our CPNA can possibly be used to provide nanolasing as well. In short, our numerical simulations show indeed that the proposed system can be used as a CPNA for the 2 divergent sources and also as a nanolaser.

## 4. Discussion and Conclusions

Both analytical dipole model analyses and numerical simulations have shown that an absorbing nanoparticle properly placed inside a DNG metamaterial slab can absorb up to 100% incoming energy from 2 identical divergent sources placed symmetrically outside of a slab. This allows us to put forward the concept of CPNA for diverging beams. Furthermore, in contrary to CPA with usual dielectric cavity [1-5], the CPNA device proposed is robust to small perturbations in the properties of the absorbing nanoparticle and the metamaterial as well as the position of the sources.

The generalization of this approach to 3D systems is straightforward because it has been shown that in a 3D system of the sources and sinks, analogous concentration and absorption of energy are also possible [10,11]. Besides, if we put our system in a properly tuned cavity, we can provide 100% absorption of energy emitted by 2 sources.

Importantly, our approach can also be applied to create single side perfect nanoabsorbers. Indeed due to the symmetry of our solution, one can simply cut the system along the symmetry plane and put a perfect electric conductor (PEC) or perfect magnetic conductor (PMC) mirror here (see Fig.9). Furthermore, our concept can be generalized to more complicated geometries such as wedge and sphere.

Let us stress again that our system allows focusing radiation in nanoscale regions and can be applied in optical nanodevices of different purposes. For example, one may use double side or single side CPNA to arrange readout of the results of quantum computation which are based on single photon qubits.

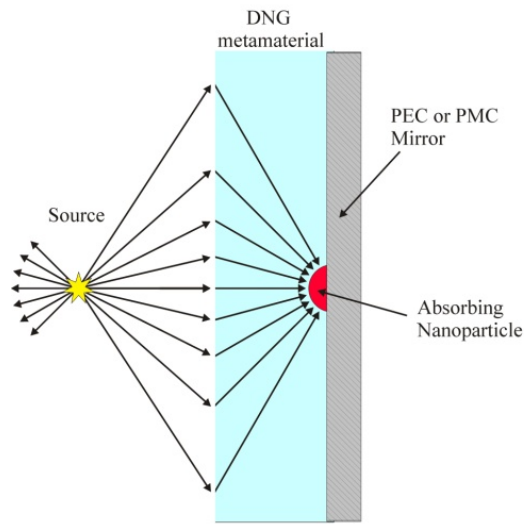

Fig. 9. Schematic diagram of a single-side coherent perfect nanoabsorber. A perfect electric conductor (PEC) [perfect magnetic conductor (PMC)] mirror should be used for the electrically [magnetically] polarized source.

**Acknowledgments**

V. Klimov would like to express his gratitude to the Russian Foundation for Basic Research (grants ## 11-02-91065, 11-02-92002, 11-02-01272, 12-02-90014, 12-02-90417) and the Presidium of the Russian Academy of Sciences for financial supports. S. Sun and G. Y. Guo would like to thank the National Science Council and National Center for Theoretical Sciences of Taiwan as well as Center for Quantum Science and Engineering, National Taiwan University (CQSE-10R1004021) for financial supports. The authors thank Konstantin Simovski and Olivier Martin for fruitful discussions and valuable comments on this work.